\documentclass{elsart}

\usepackage{graphicx}

\usepackage{amssymb}

\newcommand{\ie}{{\em i.e.\/}}
\newcommand{\eg}{{\em e.g.\/}}

\newcommand{\vlj}{V_\mathrm{LJ}}
\newcommand{\vljs}{\vlj^\mathrm{shifted}}
\newcommand{\vfene}{V_\mathrm{FENE}}
\newcommand{\vba}{V_\mathrm{BA}}
\newcommand{\vsh}{V_\mathrm{SH}}
\newcommand{\vst}{V_\mathrm{ST}}
\newcommand{\rcut}{r_\mathrm{cutoff}}

\begin{document}

\begin{frontmatter}



\title{A Simple Computer Model for Liquid Lipid Bilayers}


\author{Olaf~Lenz\corauthref{cor1}}\ead{olenz@Physik.Uni-Bielefeld.DE}, 
\author{Friederike~Schmid}\ead{schmid@Physik.Uni-Bielefeld.DE}

\corauth[cor1]{Corresponding author.}

\address{Condensed Matter Theory, Fakult\"at f\"ur Physik,
  Universit\"at Bielefeld, Postfach~100131, D-33501~Bielefeld,
  Germany}

\begin{abstract}
  We present a simple coarse-grained bead-and-spring model for lipid
  bilayers. The system has been developed to reproduce the main
  (gel--liquid) transition of biological membranes on intermediate
  length scales of a couple of nanometres and is very efficient from a
  computational point of view. For the solvent environment, two
  different models are proposed. The first model forces the lipids to
  form bilayers by confining their heads in two parallel planes. In
  the second model, the bilayer is stabilised by a surrounding gas of
  ``phantom'' solvent beads, which do not interact with each
  other. This model takes only slightly more computing time than the
  first one, while retaining the full membrane flexibility. We
  calculate the liquid--gel phase boundaries for both models and find
  that they are very similar.
\end{abstract}

\end{frontmatter}

\section{Introduction}
\label{sec:introduction}

A biological membrane (\eg{} the cell wall or walls between different
compartments in the cell) consists mainly of a bilayer of amphiphilic
lipids.  It forms an impermeable barrier between watery environments
that may have different chemical characteristics and it is extremely
flexible.  This is generally believed to be due to the \emph{liquid}
state of the bilayer. In this state, the hydrophobic tails of the
lipids are disordered and the whole bilayer shows a very low viscosity
in the membrane plane \cite{BEM91}.

However, for many lipids the main transition -- where the bilayer
undergoes a phase transition to the \textit{gel} phase with a much
higher viscosity and ordered tails -- is close to body temperature
(\eg{} $\approx 41 ^oC$ in a pure DPPC bilayer).  It seems reasonable
to assume that this proximity of the transition carries some
biological importance, \eg{} for lipid-mediated protein-protein
interactions or for the formation of rafts and domains on the
nanometre scale.  Therefore it is interesting to study the properties
of the transition.

The liquid--gel transition of the bilayer is mainly characterised by a
change of the in-plane viscosity, the area of the bilayer per lipid
and the lipid tail ordering.  These effects take place on an
intermediate length scale of a few up to a few tens of nanometres and
involve some ten thousands of atoms.  Only recently it has become
possible to access these length scales by experimental techniques
(\cite{MJ97,He2000}).

Computer simulations have proved to be excellent tools for
understanding the nature of phase transitions in this type of systems.
However, on current computer hardware, systems of a size of some ten
thousands of atoms can not be studied in detail by simulations on the
atomic level.  To bridge the gap between the length scales, we propose
a simple, robust, coarse-grained model of a lipid bilayer. It is
devised to reproduce the liquid--gel transition while spending very
little computing time the solvent environment. The model does not
attempt to catch the chemical details of any specific lipid, but
instead describes the generic characteristics of a bilayer that
undergoes the main transition.

Comparable models have been used before by G\"otz and Lipowsky
\cite{GL98}, Sintes and Baumg\"artner \cite{SB97,SB98a,SB98b}, Noguchi
and Takasu \cite{Nog2002,NT2001a,NT2001b,NT2002a,NT2002b} and Farago
\cite{Fa2003}.  Of these models, only the latter has been shown to
exhibit the liquid--gel phase transition. Otherwise, the models differ
mainly in the treatment of the solvent environment. A good model of
the solvent environment is required to keep the bilayer stable in the
liquid phase. G\"otz and Lipowsky employ a rather costly model of
explicit solvent beads, whereas Sintes and Baumg\"artner use a surface
potential similar to the surface potential solvent model proposed in
this article. The models of Noguchi and Takasu and of Farago both are
solvent-free, \ie{} they do not use an explicit solvent model, which
makes them very efficient. Instead, they use elaborate lipid--lipid
potentials to acquire liquid bilayer stability.  The phantom bead
solvent model described in this article is as efficient as these
solvent-free models. However, it is simpler and presumably more
robust. The phantom beads have a simple physical interpretation.

The model has been tested using a constant-pressure Monte-Carlo (MC)
simulation, but it should also be possible to use it in Molecular
Dynamics (MD) computer simulations.

\section{Model}
\label{sec:model}

The bilayer model consists of two subparts -- the model of the lipids
that form the bilayer (Section \ref{sec:lipid_model}), and the solvent
environment model (Section \ref{sec:solvent_model}). The solvent
environment is required to keep the liquid bilayer together.  The model
is robust towards different solvent models in the sense that the exact
form of the model solvent seems to be relatively unimportant. In the
following, we will present two very different solvent environment
models, both of which enable the bilayer to undergo the liquid--gel
transition and maintain a stable liquid phase, while still being
efficient when it comes to computing.

\subsection{Lipid Model}
\label{sec:lipid_model}

The lipid model used in this work was derived from a successful
coarse-grained model for Langmuir monolayers that was able to
reproduce the generic phase diagram of such monolayers in great detail
\cite{SLS99,DS2001}.

\begin{figure}[htbp]
  \centering
  \includegraphics{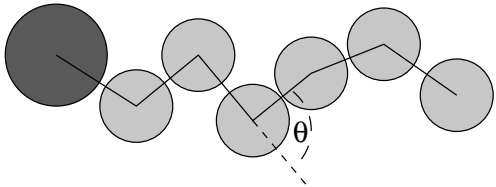}
  \caption{Sketch of a lipid.}
  \label{fig:lipid_model}
\end{figure}
A lipid in the model system consists of six tail beads and one
slightly larger head bead (see Figure \ref{fig:lipid_model}).  
All lipid beads interact via a truncated 12-6-Lennard-Jones potential
of the form :
\begin{equation}
  \label{eq:vlj1}
   \vljs (r) = \left\{ 
    \begin{array}{ll}
      0 & \mathrm{, if~} r > \rcut\\
      \vlj (r) - \vlj (\rcut) & \mathrm{, otherwise}\\
    \end{array}
  \right.
\end{equation}
with
\begin{equation}
  \label{eq:vlj}
  \vlj (r) = \epsilon \left(
    {\left( \frac{\sigma}{r} \right)}^{12} 
    - 2 {\left( \frac{\sigma}{r} \right)}^6 \right)
\end{equation}

For tail-tail interactions, the cutoff is $\rcut = 2 \sigma$, the
potential thus has an attractive contribution. Head beads interact
with each other and with tail beads via a soft-core potential, \ie{}
the purely repulsive core of the 12-6-Lennard-Jones potential $\vljs$
with a cutoff radius of $\sigma$.

The adjacent beads of a lipid chain are bound to each other by a FENE
(finite extensible nonlinear elastic, see Equation \ref{eq:vfene})
type spring potential that limits the maximal and minimal bond length.
Additionally, a bond-angle potential (Equation \ref{eq:vba}) favours
stretched chain conformations.
\begin{equation}
  \label{eq:vfene}
  \vfene (r) = - \frac{1}{2} \epsilon ( \Delta r_\mathrm{max} )^2 
  \> \log \left( 1 - \frac{r-r_0}{\Delta r_\mathrm{max}}^2 \right)
\end{equation}
\begin{equation}
  \label{eq:vba}
  \vba (\theta) = \epsilon \> (1 - \cos \theta)
\end{equation}

The length of the chains (\ie{} the number of tail beads) and the
interaction parameter of the bond-angle potential $\vba$ are adapted
to model lipids with a fully saturated acyl chain of 16 to 18 carboxyl
groups \cite{Sta98,SLS99,DS2001}. The interaction potentials and
parameters of the lipid model are summarised in table
\ref{tab:potentials}. All lengths in the system are measured in units
of the tail--tail-potential Lennard-Jones parameter $\sigma$, the
energy unit is $k_B T$.

\begin{table}[htbp]
  \centering
  \begin{tabular}{l|l|l}
    \textbf{} & \textbf{Potential} & \textbf{Parameters}\\\hline
    tail--tail & $\vljs (|\vec{r}|)$ & $\epsilon = 1$, $\sigma = 1$, $\frac{\rcut}{\sigma} = 2$\\
    head--tail & $\vljs (|\vec{r}|)$ & $\epsilon = 1$, $\sigma = 1.05$, $\frac{\rcut}{\sigma} = 1$\\
    head--head & $\vljs (|\vec{r}|)$ & $\epsilon = 1$, $\sigma = 1.1$, $\frac{\rcut}{\sigma} = 1$\\
    bond-length & $\vfene (|\vec{r}|)$ & $\epsilon = 100$, $r_0 = 0.7$, $\Delta
    r_\mathrm{max} = 0.2$\\
    bond-angle & $\vba (\theta)$ & $\epsilon = 4.7$\\
  \end{tabular}
  \caption{Summary of the interaction potentials of the lipid
    model. $\vec{r}$ is the centre-to-centre distance vector of two
    beads, $\theta$ is the bond angle of three adjacent beads (as
    depicted in figure \ref{fig:lipid_model}). Parameters are taken
    from \cite{DS2001}.}
  \label{tab:potentials}
\end{table}

\subsection{Solvent Environment Models}
\label{sec:solvent_model}

In the first solvent environment model used in this work, the bilayer
is defined by two parallel planes. The lower plane is the
$x$-$y$-plane itself, the upper plane is shifted by $z_\mathrm{upper}
> 0$. The tail beads of the bilayer are confined between the planes by
the surface potential $\vst$ (Equation \ref{eq:surface_tail}), while
the head beads are forced to stay above the upper plane resp. below
the lower plane by $\vsh$ (see Equation \ref{eq:surface_head}, with
the parameters $\epsilon=10$, $r_0=0$ and $\Delta r_\mathrm{max}=0.5$
for $\vfene$). 

\begin{equation}
  \label{eq:surface_tail}
   \vst (r) = \left\{ 
    \begin{array}{ll}
      \vfene(z) & \mathrm{, if~} z < 0 \\
      \vfene(z-z_\mathrm{upper}) & \mathrm{, if~} z > z_\mathrm{upper}\\
      0 & \mathrm{, otherwise}\\
    \end{array}
  \right.
\end{equation}

\begin{equation}
  \label{eq:surface_head}
   \vsh (z) = \left\{ 
    \begin{array}{ll}
      \vfene(z) &
      \mathrm{, if~} 0 < z < \frac{1}{2} z_\mathrm{upper} \\
      \vfene(z-z_\mathrm{upper}) & 
      \mathrm{, if~} \frac{1}{2} z_\mathrm{upper} < z < z_\mathrm{upper}\\
      0 & \mathrm{, otherwise}\\
    \end{array}
  \right.
\end{equation}

Unfortunately, the bilayer is not flexible, so the model is useless
for the simulation of phenomena that involve any membrane
deformations, such as undulations, hydrophobic mismatch effects of
membrane integral proteins etc.

On the other hand, this solvent environment model is very easy to
implement and is also very efficient when it comes to computing, as it
will add only a single term per bead to the energy sum.

Therefore, another solvent environment model was developed, which
retains the full membrane flexibility, while adding only a small
computational overhead. In this ``phantom solvent bead model'', the
solvent is represented by explicit solvent beads. These beads behave
exactly like additional, unbound head beads (\ie it has a purely
repulsive soft-core interaction with the lipid beads), except that they do
not interact with each other.

This has a number of advantages. First, the model is still
computationally efficient, as only those solvent beads that are
actually close to the bilayer significantly contribute to the
computing time. Furthermore, the model has no solvent artefacts, as
the solvent can not develop any internal structure. Therefore, the
model can employ periodic boundary conditions perpendicular to the
bilayer and only a relatively thin layer of solvent beads is needed to
ensure that the bilayer does not interact with itself via the periodic
boundary conditions. If one uses a thicker layer of phantom solvent
beads, the pressure of the system can be easily measured, as the ideal
gas equation of state holds for the phantom solvent beads far from the
bilayer.

\section{Simulation Details}
\label{sec:details}

The system was simulated using a constant pressure Monte-Carlo (MC)
simulation with periodic boundary conditions in all directions.  We
used standard single bead moves as well as collective volume moves
that stretched or squeezed the whole system in $x$-, $y$- or
$z$-direction independently, the effective Hamiltonian being
\begin{equation}
  \label{eq:hamiltonian}
  H_\mathrm{eff} = \Phi(r) + p V - N k_B T \log(V)
\end{equation}
where $\Phi(r)$ is the potential energy of all beads, $p$ is the
desired pressure, $V$ is the (fluctuating) volume, $T$ is the
temperature of the system and $N$ is the number of beads in the
system. $H_\mathrm{eff}$ results from a Laplace transformation of the
$NVT$-Hamiltonian.

We note, that in the case of the surface potential solvent model, the
volume of the system $V$ is, strictly speaking, not well-defined. 
However, this does not cause problems, because $V$ in the effective
Hamiltonian just sets the length scales of the system in the different
directions.  Therefore, we can rewrite the effective Hamiltonian for
the surface potential model as
\begin{equation}
  H_\mathrm{eff}^\mathrm{surface} = \Phi(r) + p L_x L_y
  z_\mathrm{upper} - N k_B T \log(L_x L_y
  z_\mathrm{upper})
\end{equation}
where $L_x$ and $L_y$ are the system sizes in $x$ and $y$-direction
respectively \cite[pages 103ff]{FS96}.

The simulation runs performed for this work involved 288 lipids, which
were initially set up perpendicular to the $x$-$y$-plane with 144
lipids in each layer, the heads forming a two-dimensional triangular
grid. In the case of the phantom solvent bead model, up to 16 666
solvent beads were distributed randomly outside the bilayer.

The total run length varied between 500 000 and 2 000 000 Monte-Carlo
steps (MCS), where one MCS includes one volume move in each direction
and $N$ single bead moves. The maximal move ranges for the different
moves were adapted to yield an acceptance rate of approximately 30 \%
in a simulation pre-run of 10 000 to 50 000 MCS.

\section{Results}
\label{sec:results}

\subsection{Liquid--gel phase transition}

\begin{figure}[htbp]
  \centering
  \includegraphics{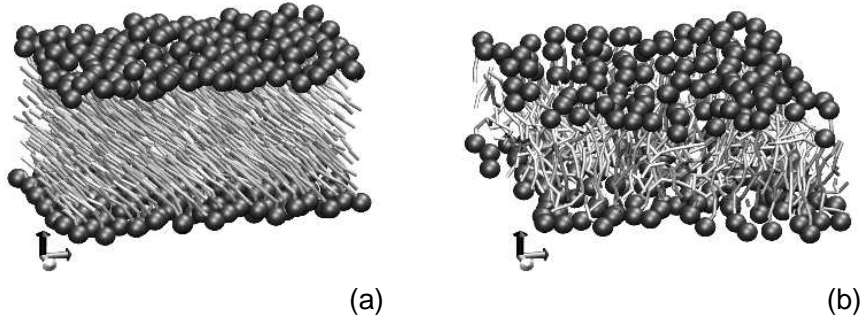}
  \caption{Snapshots of the bilayer model with surface potential
    in the gel ((a), $p=1.0$, $T=0.9$) and liquid
    ((b), $p=1.0$, $T=1.0$) phase. The tail
    beads have been replaced by tubes for better visualisation.}
  \label{fig:conf_surface}
\end{figure}

\begin{figure}[htbp]
  \centering
  \includegraphics{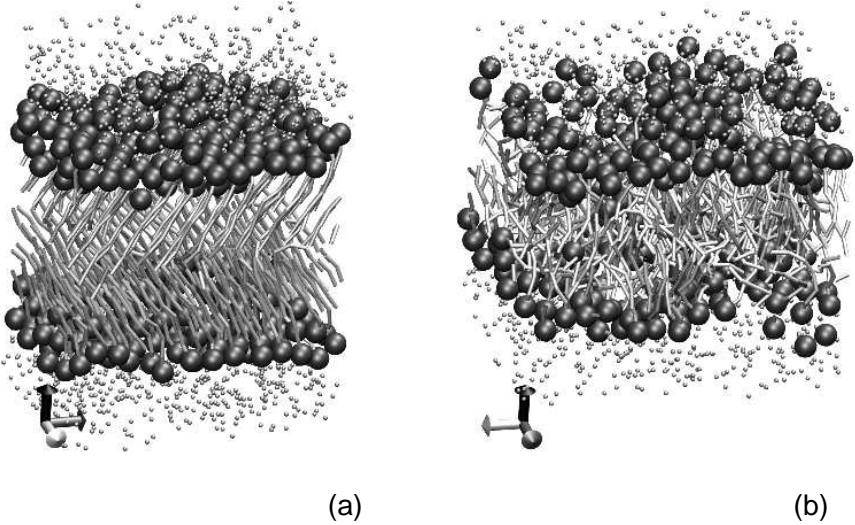}
  \caption{Snapshots of the bilayer model with phantom solvent beads
    in the gel ((a), $p=1.0$, $T=0.9$) and liquid
    ((b), $p=1.0$, $T=1.0$) phase. The tail
    beads have been replaced by tubes and the phantom solvent beads by
    small spheres for better visualisation.}
  \label{fig:conf_phantom}
\end{figure}

Figures \ref{fig:conf_surface} and \ref{fig:conf_phantom} show typical
snapshots of systems with both solvent environment models in the gel
and liquid phases. The gel phase is characterised by a strong ordering
the lipid tails, a high bilayer thickness and a small area per lipid
of the bilayer, while the liquid phase exhibits disordered tails, a
much lower bilayer thickness and a larger area per lipid head.

\begin{figure}[htbp]
  \centering
  \includegraphics{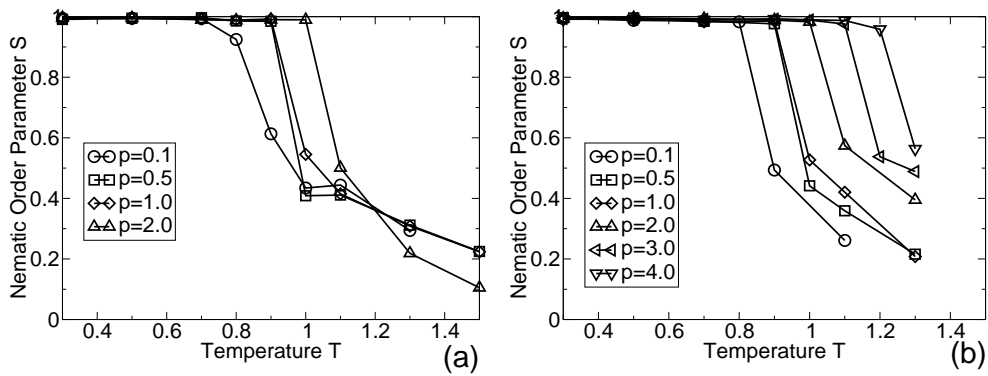}
  \caption{Plot of the nematic order parameter $S$ of the system with
    surface potential (a) and phantom solvent beads (b).}
  \label{fig:S}
\end{figure}

\begin{figure}[htbp]
  \centering
  \includegraphics{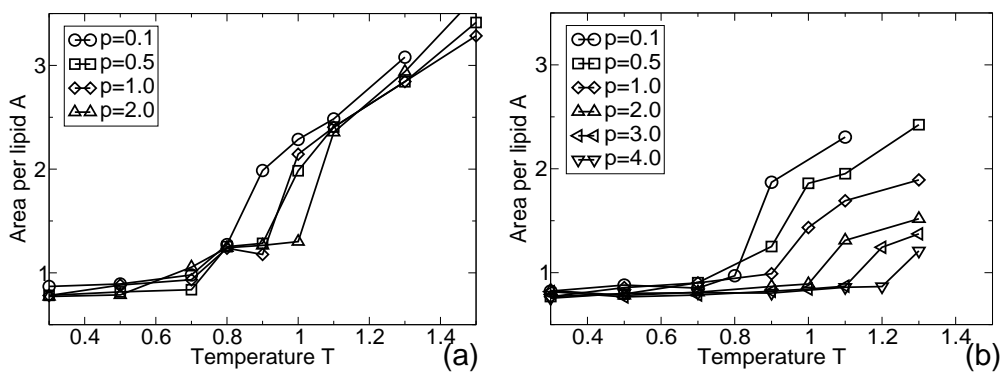}
  \caption{Plot of the area per lipid A of the system with
    surface potential (a) and phantom solvent beads (b).}
  \label{fig:A}
\end{figure}

The lipid tail ordering can be expressed by the nematic order
parameter $S$ of the lipid chains which is given by the largest
eigenvalue of the matrix
\begin{equation}
  \label{eq:S}
  S_{ij} = \frac{1}{2N} \sum_{n=1}^{N_\mathrm{lipids}} ( 3 x_i^{(n)} x_j^{(n)} - \delta_{ij})
\end{equation}
where $x_i$ and $x_j$ are the components of the end-to-end vectors
of the $N_\mathrm{lipids}$ lipid chains in a configuration \cite{GP93}.
As shown in Figure \ref{fig:S}, it drops considerably at the phase
transition.  Figure \ref{fig:A} displays the jump of the measured area
per lipid $A$ at the phase transition.

\subsection{Phase Diagrams}

\begin{figure}[htbp]
  \centering
  \includegraphics{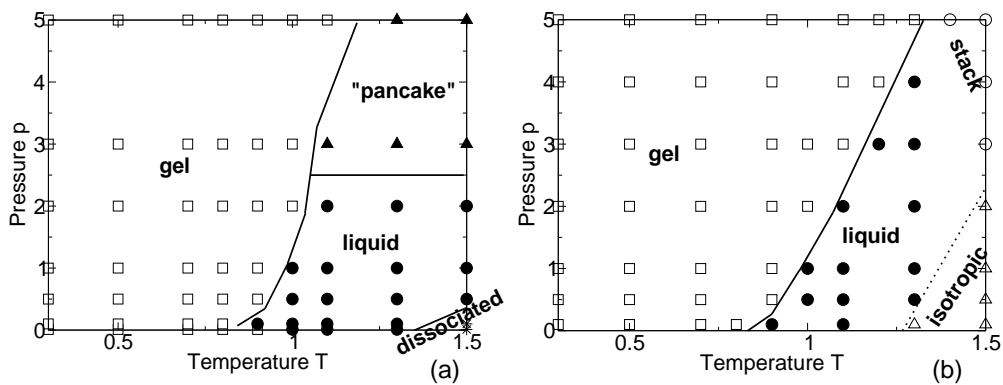}
  \caption{Phase diagrams of the bilayer model with surface potential
  (a) and with phantom solvent beads (b). The symbols indicate
  parameter values for which simulations have been carried out. Open
  squares: gel phase, closed circles: liquid phase, closed triangles:
  ``pancake'' phase, stars: dissociated monolayers, open circles:
  spontaneous formation of bilayer stacks, open triangles: disordered
  isotropic phase. The lines are eye-guides only and indicated the
  approximate location of the phase boundaries.}
  \label{fig:pd}
\end{figure}

Figure \ref{fig:pd} depicts the phase diagrams of the bilayer model
obtained from the simulations with both solvent environment models.
The liquid--gel transition can be observed clearly in both systems at
low to intermediate pressures.

At higher temperature and lower pressure, the bilayer breaks up into
two dissociated monolayers.  At higher pressures, the surface
potential model develops an unphysical ``pancake'' phase. In this
phase, the distance of the two layers ($z_\mathrm{upper}$) almost
becomes zero and the lipids spread over a large area, as in a
two-dimensional gas.

In the phantom solvent bead model, the liquid--gel phase transition
can be observed over a wider pressure range. At lower pressure and
higher temperature, the liquid bilayer dissolves and forms an
isotropic phase.  At higher pressure and higher temperature, we have
also observed the spontaneous formation of a stack of two liquid
bilayers.

\section{Conclusions and Discussion}
\label{sec:discussion}

From the Langmuir monolayer model used in our group before
\cite{DS2001,SS99,SLS99} we learned, that the key to the successful
reproduction of the generic phase behaviour of lipid systems is the
qualitatively correct modelling of the translational and
conformational degrees of freedom of the lipids.  Furthermore, the
attractive part of the tail-tail potential is crucial for the liquid
phase and the liquid--gel transition.  As long as this is taken into
account, however, the liquid--gel transition is robust towards changes
in the exact form and parameters of the interaction potentials between
the lipids.  It is to be expected that these results also hold for the
bilayer model.

As far as the solvent environment model is concerned, we can deduce
that the existence and location of the transition is almost not
influenced by the solvent model. In contrast to this, the stability of
the liquid phase does depend on the solvent.  However, it seems to be
sufficient to introduce ``phantom beads'' that basically probe the
free solvent-accessible volume to create a stable, liquid
bilayer.

Furthermore, the model is very efficient with both proposed solvent
environment models. Even in the phantom solvent model only those
solvent beads significantly contribute to the computing time that
actually interact with the lipids.

We conclude that the presented lipid model in combination with a
simple solvent environment model is well suited for simulating lipid
bilayers in the regime of the liquid--gel phase transition.  The model
correctly describes the strong decrease of the lipid tail ordering and
bilayer thickness, as well as the increase of the area per lipid.

The phantom solvent model retains the full flexibility of the liquid
bilayer.  This is necessary, as the model will be used to investigate
lipid-mediated interactions between membrane integral proteins.  The
model presented here is ideally suited for this task.

\section{Acknowledgements}
\label{sec:ack}

The configuration snapshots have been created using VMD\cite{vmd}.  

We thank Harald Lange for the implementation of the simulation
prototype, Dominik D\"uchs for the Langmuir monolayer code and Claire
Loison for useful discussions. The genetics group of the University of
Bielefeld has kindly provided us with additional computing time. This
work was funded by the German Science Foundation (DFG) within the
collaborative research centre SFB 613.

\bibliography{paper}
\bibliographystyle{elsart-num}

\end{document}